\documentclass{article}

\usepackage{tikz}
\usetikzlibrary{positioning, arrows.meta}
\usepackage{tcolorbox}
\tcbuselibrary{breakable}

\usepackage[a4paper,margin=25mm, textwidth=12cm]{geometry}
\usepackage{blindtext,setspace,amsmath,graphicx,hyperref,xcolor,amssymb}
\title{\textbf{Opinion Consensus Formation Among Networked} \\ \textbf{Large Language Models}}

\author{
Iris Yazici\\
Bilkent University\\
\and
Mert Kayaalp\\
UBS-IDSIA AI Lab 
\\
\and
Stefan Taga\\
EPFL\\
\and
Ali H. Sayed\\
EPFL\\
}

\date{} 

\begin{document}

\maketitle
\vspace{-3.5em} 

\vspace{1.5em}

\begin{abstract}
Can classical consensus models predict the group behavior of large language models (LLMs)? We examine multi-round interactions among LLM agents through the DeGroot framework, where agents exchange text-based messages over diverse communication graphs. To track opinion evolution, we map each message to an opinion score via sentiment analysis. We find that agents typically reach consensus and the disagreement between the agents decays exponentially. However, the limiting opinion departs from DeGroot's network-centrality-weighted forecast. The consensus between LLM agents turns out to be largely insensitive to initial conditions and instead depends strongly on the discussion subject and inherent biases. Nevertheless, transient dynamics align with classical graph theory and the convergence rate of opinions is closely related to the second-largest eigenvalue of the graph’s combination matrix. Together, these findings can be useful for LLM-driven social-network simulations and the design of resource-efficient multi-agent LLM applications.
\end{abstract}
\vspace{1em}

\section{Introduction}
\label{sec:intro}

Mathematical models of opinion formation have been instrumental in understanding social network behavior and in designing multi-agent information-processing systems. The DeGroot consensus model \cite{degroot1974}, in particular, has received a lot of attention from the signal processing, microeconomics, and control communities, where many variants \cite{sayed2014,krishnamurthy_2013, djuric2012, inan2022social, salhab2020cdc,kayaalp2022random} have been proposed to better reflect group decision dynamics and to design better decentralized signal processing and optimization algorithms.

In this work, we study social interactions among large language models (LLMs) through the lens of the DeGroot framework. Our goals are twofold. 
First, despite a rich theoretical literature, empirical validation in real-world social settings remains limited, largely because collecting behavioral data from human subjects is costly and time-consuming.
We examine whether LLMs can provide a controllable and low-cost testbed for modeling the dynamics observed in human social networks.
Second, as LLMs are increasingly deployed in real applications, including social media platforms, it becomes important to understand how opinions evolve within networks of LLM-based agents.

To that end, we conduct experiments where LLM agents interact on communication graphs, and exchange text-based statements with their immediate neighbors over multiple rounds. We enforce the network weights and agent characters through system prompts, conduct multi-round simulations, and assign opinion scores to all responses via sentiment analysis. We vary the topics and graph topology, and compile the resulting conversations into a dataset. We then analyze the underlying opinion dynamics.

We find that agents typically converge to a consensus. Somewhat surprisingly, however, the final beliefs are largely insensitive to their initial positions, which is a departure from the DeGroot prediction that consensus should equal a network-centrality–weighted average of initial opinions. Instead, the consensus point appears to be dependent on the discussion subject and on biases the LLM carries, possibly from pretraining and alignment phases of their training. Despite this mismatch, the rate of convergence to the consensus agrees with well-established graph-theoretical results: it is related to the second-largest modulus eigenvalue of the combination matrix \cite{jackson2008social,xiao2003fast}. Moreover, we find that when the combination matrix's weights are instructed to the LLM agents through system prompts, agents are more likely to reach a consensus. In order to stimulate further research, we open-source our dataset totaling 764 experiments with 8 different topics, containing more than 1,200,000 LLM responses. It is available on Hugging Face \footnote{https://huggingface.co/datasets/asl-epfl/Social-LLM-Networks}.  

\subsection{Related Work}

Models of opinion dynamics and distributed inference study how a network of agents aggregates and updates its beliefs by repeatedly incorporating others’ views. The standard DeGroot model \cite{degroot1974} captures this via repeated averaging, where each agent forms a weighted average of neighbors’ opinions over time. Beyond DeGroot, richer variants allow for Bayesian reasoning at the agent level \cite{acemoglu2011opinion} and for distinct communication patterns, such as private interactions \cite{inan2022social} or randomized processes \cite{salhab2020cdc,kayaalp2022random}. There is also experimental work evaluating these mechanisms in real-world social settings \cite{chandrasekhar2020testing, mobius2015treasure, mueller2013social}, though such empirical tests remain relatively rare.

Because experiments within human social networks are difficult to run, recent work examines the use of LLMs for simulation. They have been used for political processes \cite{ferraro2024}, social-platform design choices \cite{tornberg2023}, spread of misinformation \cite{li2024}, and for replicating classical social-psychology experiments \cite{borah2025mind}. Motivated by the emergence of human-LLM collectives, there is also a line of work that frames LLMs as ``distributed sensor networks'' integrating textual inputs \cite{jain2025interacting}.

Likewise, opinion dynamics among LLMs have been studied to assess how closely they mirror social networks. For example, \cite{chuang2023simulating} finds that standard LLMs initialized with human-like personas tend to reach factual consensus, likely due to their knowledge priors. Similarly, \cite{cisneros-velarde2025} shows intrinsic training-induced biases: alignment-trained models exhibit a consensus-seeking tendency even when used as-is, without persona prompts. In comparison, we analyze opinion dynamics from a DeGroot perspective, and focus on topics that are also debated in real-world social networks rather than factual questions with ground-truth answers. Furthermore, we study the impact of the communication network topology by evaluating the consensus rate among LLMs. Doing so assists in clarifying the potential of LLMs for social-simulation research, and also helps guide the design of multi-agent LLM systems that are resource-efficient in both communication and context-length.

\section{Problem Formulation}
\label{problem_formulation}

We consider a network of $K$ agents. 
Initially, each agent $k$ has a belief vector (i.e., opinion) $\mu_{k,0}$, which it updates based on the opinions of its neighbors. 
The peer-to-peer communication is constrained on a weighted and directed graph topology. Each agent $k$ receives information from its neighbors $\mathcal{N}_k$. In the DeGroot model \cite{degroot1974}, agents repeatedly average their neighbors' opinions in order to update theirs. Namely, at each time instant $i$, agent $k$ updates its belief with 
\begin{equation}\label{eq:linear_update}
    \mu_{k,i} = \sum_{\ell \in \mathcal{N}_k} a_{\ell k} \mu_{\ell, i-1}, 
\end{equation}
where $a_{\ell k}$ denotes the level of trust agent $k$ assigns to the belief vector it receives from agent $\ell$. These coefficients satisfy
\begin{equation}
    \sum_{\ell \in \mathcal{N}_k} \! \! a_{\ell k} = 1 \ \ \text{and} \ \  a_{\ell k}>0 \ \text{if, and only if,} \ \ell \in \mathcal{N}_k, \ \text{0 if} \ \ell \notin \mathcal{N}_k
\end{equation}
where the combination matrix $A = [a_{\ell k}]$ is left-stochastic. If the underlying graph is also strongly connected \cite{sayed2014}, that is, if there exists a path between any agent pair $(\ell,k)$ and there exists at least one agent $k$ that does not discard its own information (i.e., $\exists k$ $a_{kk} > 0$), then, the matrix $A$ becomes both aperiodic and irreducible. This implies that under \eqref{eq:linear_update} and by the Perron-Frobenius theorem \cite{sayed2014, pillai2005perron}, all agents will reach consensus with asymptotic beliefs given by 
\begin{equation}\label{eq:limit_consensus}
    \lim_{i \to \infty} \mu_{k,i} = \sum_{\ell = 1}^K \pi_{\ell} \  \mu_{\ell, 0}
\end{equation}
Here, $\pi$ denotes the Perron eigenvector of the combination matrix $A$ and satisfies 
\begin{equation}
    A \pi = \pi , \ \ \  \sum_{k = 1}^K \pi_{k} = 1, \ \   \text{and} \ \forall k, \ \pi_{k} > 0.
\end{equation}
Entry $\pi_k$ represents how central agent $k$ is in the network. Equation \eqref{eq:limit_consensus} then means that the final opinions of all agents are going to be the same and equal to a weighted average of initial opinions. 

A strongly connected graph also implies that the matrix $A$ has a unique eigenvalue at $1$ and all its other eigenvalues are strictly smaller than $1$ in absolute value. 
It is also known that the second-largest magnitude eigenvalue $|\lambda_2|< 1$ of $A$ controls the convergence time.
This follows from \cite[Chapter 8]{horn2012matrix}
\begin{equation}\label{eq:horn_convergence}
    \left | [A^t]_{\ell k} - \pi_{\ell} \right | \leq C_{\sigma} \cdot \sigma^t,
\end{equation}
which holds for any $\sigma$ that satisfies $|\lambda_2| < \sigma < 1$ and for some constant $C_{\sigma}$ that does not depend on $t$.
Therefore, the convergence to consensus is exponentially fast and the rate of convergence is inversely proportional to $\lambda_2$.
Note that, in general, $\lambda_2$ decreases with increasing network connectivity, and attains its minimum value at $0$ if, and only if, the underlying graph is fully-connected with a rank-one combination matrix. 

\subsection{Networked LLM agents}

In this work, we evaluate LLM-based agents under the DeGroot framework. 
For the graph topology, we use Erd\H{o}s--R\'enyi random graphs with varying connectivity parameter $p$. In the Erd\H{o}s--R\'enyi graph model $G(K,p)$, each possible directed edge between a pair of nodes is included independently with probability $p$. To ensure the networks constructed in this manner are highly likely to be connected, and also to avoid trivial cases with isolated agents, we use the following lower bound for $p$:
\begin{equation}\label{eq:connectivity_threshold}
    p^\star \triangleq \dfrac{\ln{K}}{K}, 
\end{equation}
where $p^\star$ is known to be the connectivity threshold. Specifically, an Erd\H{o}s--R\'enyi random graph is known to be connected with high probability if $p > p^\star$ as $K \to \infty$ \cite{bollobas1998}. 
In addition to Erd\H{o}s--R\'enyi random graphs, we also consider two extreme cases, a fully connected and a circular graph topology, to better examine the effect of network connectivity.

As in the DeGroot framework, communication networks are directed, weighted. The combination matrix's weights (i.e., interaction weights and self-weights) are enforced through system prompts, which are provided to the LLMs repeatedly after each interaction round. Note that during an experiment, the graph combination matrix is held constant.
We create two kinds of agent characters depending on the self-weights:
\textit{Self-confident} agents are instructed to rely more on their own previous opinions in comparison to \textit{open-minded} agents. 
The remaining weights, that is, the portion that is not assigned to the individual self-weights, is then distributed equally among the agent's neighbors, as specified by the system prompt. 

During the interactions, in addition to this system prompt, an agent's own previous response and its neighboring agents' responses are provided to it so that it can form its new opinion accordingly.
Moreover, during the initialization phase, each LLM agent starts with an initial opinion over a topic of discussion, which is also a debatable topic between humans over real social networks. 
Agents are initially either \textit{for}, \textit{neutral}, or \textit{against} the topic, and this stance is enforced through the initial prompt. 

For our experiments, we utilize AutoGen \cite{wu2024autogen}, which is a programming framework for multi-agent applications. 
To map LLM agents’ text responses to beliefs (i.e., opinion scores), we employ a separate LLM, independent of the agents’ interactions, to perform sentiment analysis. 

\section{Experiments}

\subsection{Implementation details}

We now detail the settings used for the variables introduced above. We set $K=20$ and run $80$ interaction rounds, which we have empirically found to be sufficient for convergence.
We employ Google’s \texttt{Gemini 2.0 Flash} for conversation generation due to its cost efficiency and speed.

Initial opinions are drawn uniformly from \{\textit{for}, \textit{neutral}, \textit{against}\}; agent types are drawn uniformly from \{\textit{self-confident}, \textit{open-minded}\}; and discussion topics are drawn uniformly from \{\textit{Bitcoin}, \textit{Euthanasia}, \textit{Veganism}, \textit{Vaping}, \textit{Gene editing}, \textit{Ghosting}, \textit{C. Ronaldo}, \textit{Remote Work}\}.
The communication graph is selected as follows: an Erd\H{o}s--R\'enyi graph with probability $0.92$, a fully connected graph with probability $0.04$, and a circular graph (ring topology) with probability $0.04$.
For the Erd\H{o}s--R\'enyi case, $p \in \bigl(0.15,\,0.35\bigr)$ with probability $0.9$, with the lower bound chosen according to \eqref{eq:connectivity_threshold}, and 
$p \in (0.35,\,1)$ with probability $0.1$, since less connected graphs exhibit subtler convergence behavior.

All parameters, including initial and system prompts, graph topology, and agent responses are logged to a JSON file for each experiment, which can be found in the provided dataset and are used in the subsequent analyses. For sentiment analysis, we prompt OpenAI’s \texttt{gpt-5-nano} to assign an integer score in the range of $[-3,3]$, where $-3$ corresponds to ``against'' end and $3$ corresponds to ``for'' end.
These scores are also appended to the corresponding JSON files. 
For all the following analyses in this paper, we use sentiment scores after normalizing them into the interval $[0,1]$.

\subsection{Consensus across the agents}
\label{consensus_vs_time}

\begin{figure}[h]
    \centering
    \includegraphics[width=0.7\linewidth]{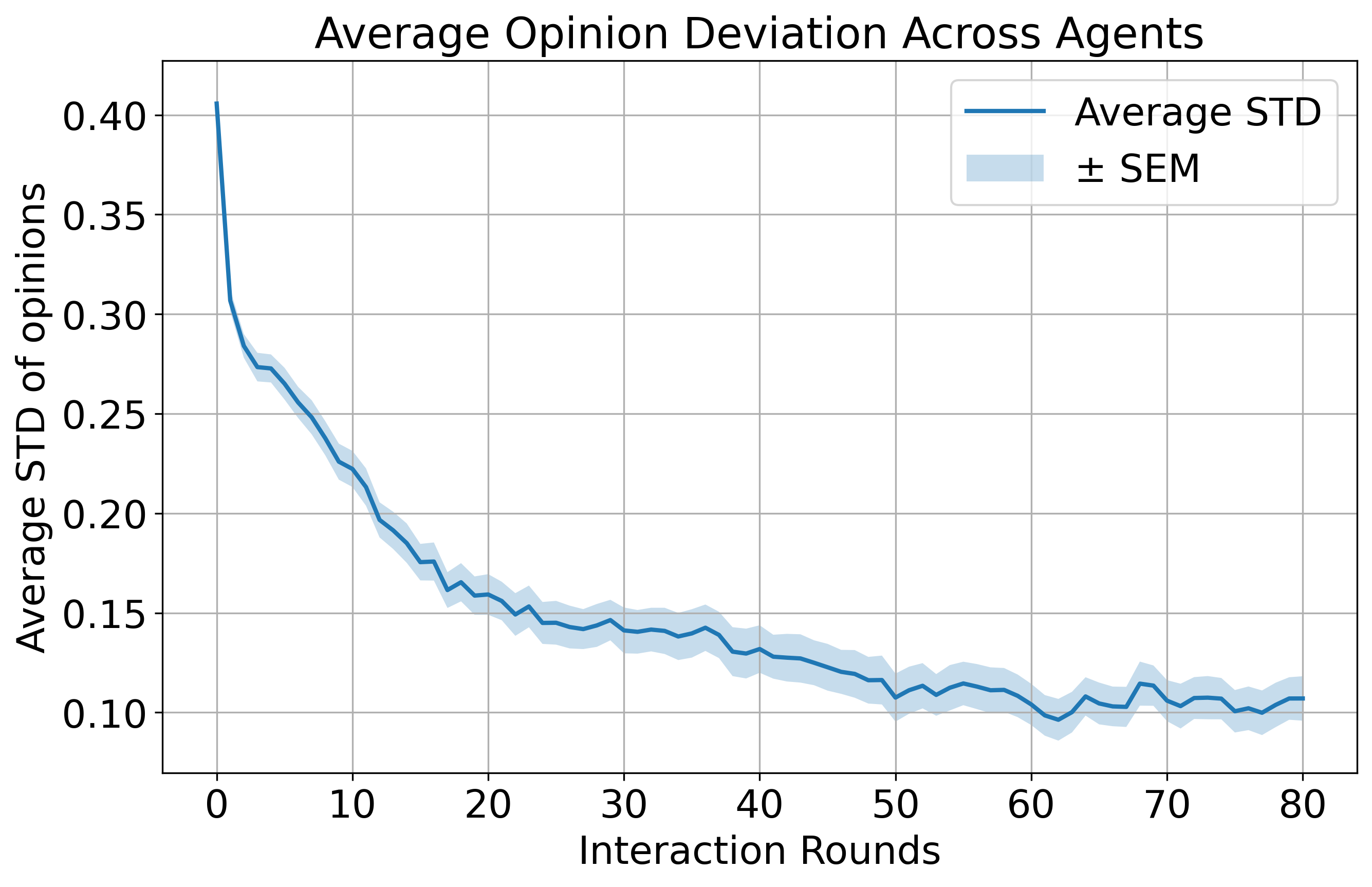}
    \caption{The average standard deviation of agents' opinions with respect to the number of iterations across $50$ bitcoin-related experiments. The shaded region indicates the standard error of the mean (SEM) across $50$ simulations.}
    \label{fig:fig_std_time}
\end{figure}

In order to investigate whether agents' reach a consensus, we conduct $315$ experiments with \textit{self-confident} and \textit{open-minded self weights} enforced through system prompts (respectively $80\%$ and $60\%$), and use the standard deviation (STD) of agents' sentiment scores as a measure of opinion deviation in each round of conversation. Figure~\ref{fig:fig_std_time} shows that, for the discussion topic of ``bitcoin'', the average STD across different experiments starts from around $0.4$, decreases exponentially (with coefficient of determination $R^2=0.965$), and approaches a steady state value close to $0.1$, demonstrating how disagreement between agents becomes negligible and a consensus is attained.

Moreover, in Table~\ref{table1}, we report the average final STD across all simulations. 
In order to reduce the effect of the noise during sentiment analysis, we take an agent's average sentiment score over the last $10$ rounds as its final opinion.
It can be seen that the average final disagreement between the agents, namely STD $= 0.083$, is almost negligible in the range of sentiment scores $[0,1]$. 
Note that in these experiments, weights were assigned through the system prompts (hence the name weighted experiments). 
We repeat the experiments by removing the self-weights assigned respectively to \textit{self-confident} and \textit{open-minded} agents in the system prompts.
Namely, we run $108$ ``weightless'' experiments, where we modify the system prompts so that even though there are still distinct \textit{self-confident} / \textit{open-minded} agents, specific self-weights are not assigned through the system prompts.
We observe that the weightless experiments exhibit a higher final diversity 
($0.165 \pm 0.008$ SEM) compared to the weighted experiments 
($0.083 \pm 0.004$ SEM). 
The difference ($\Delta = 0.082$) corresponds to $9.17$ standard errors of the 
difference ($\mathrm{SE}_\Delta = 0.00894$), yielding a $p$-value $< 0.001$ under the null hypothesis 
($H_{0}:\mu_{\text{weightless}}=\mu_{\text{weighted}}$), indicating a statistically significant increase in dispersion. 
Therefore, we conclude that enforcing weights through system prompts can increase the chances of consensus in LLM-interactions. 

\begin{table}[h]
\caption{Average final disagreement of agents. Experiment types are based on if the system prompts include instructions about self-weights.} \vspace{0.8em}
\label{table1}
\centering
\begin{tabular}{lc}
\hline
\textbf{Experiment Type} & \textbf{Average STD  $\,\pm\,$ SEM} \\
\hline
Weighted Experiments     & $0.083 \,\pm\, 0.004$ \\
Weightless Experiments & $0.165 \,\pm\, 0.008$ \\ 
\hline
\end{tabular}
\end{table}

Although we have found that agents reach a consensus, it is still not clear whether this is in line with the DeGroot consensus model.
To test the fit of the data to the DeGroot consensus, we compute the root mean squared error (RMSE) between final opinions of agents and the predicted consensus values according to \eqref{eq:limit_consensus}. 
This yields an average RMSE of $0.32$ which shows a significant mismatch between the two.
In addition, if the sentiment scores are discretized into \{\textit{for}, \textit{neutral}, \textit{against}\} categories, the average classification accuracy based on the DeGroot model is $32\%$, same as random guessing accuracy of $33\%$. 
A similar conclusion holds when the task is reduced to a binary classification between \{\textit{for}, \textit{against}\}. The accuracy here improves only to $60\%$, not significantly better than random guessing accuracy of $50\%$. 
Collectively, these results show that the DeGroot model's consensus prediction in \eqref{eq:limit_consensus} fails to capture the steady-state consensus beliefs observed in our experiments.

\subsection{Topic-dependent bias in final opinion distribution}
\label{bias}

\begin{figure}[h]
    \centering
    \includegraphics[width=0.7\linewidth]{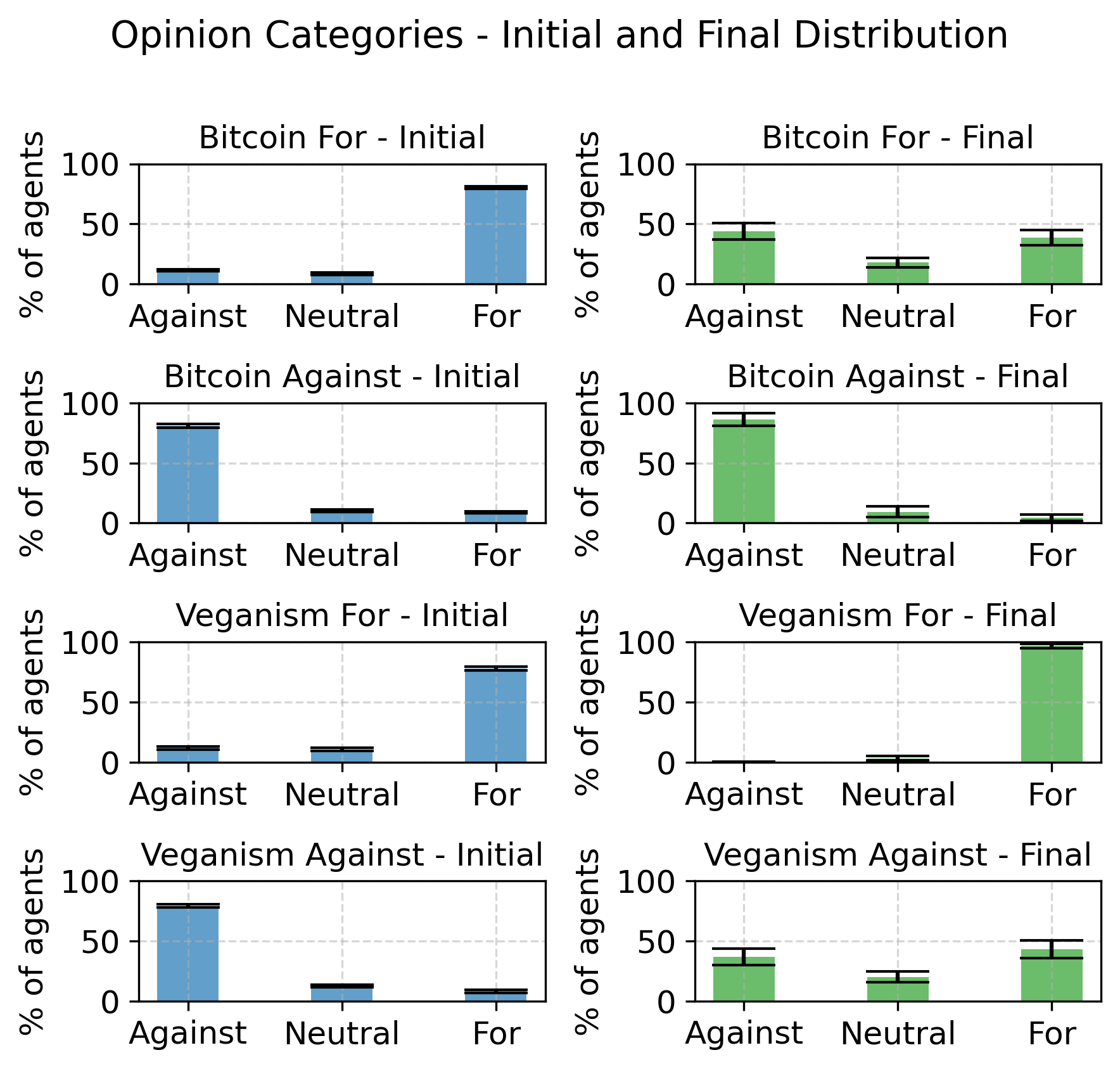}
    \caption{\textit{Left}: Initial opinion distributions, \textit{Right}: Final opinion distributions. Each row belongs to a set of experiments with a different topic and initial opinion distribution. For example, the first row denotes a set of experiments where the initial opinion distribution is highly skewed towards ``for'' on bitcoin sentiment, while for the second row, the initial majority is ``against''. The error bars denote the SEM with respect to a total of $150$ experiments.}
    \label{fig:fig_4x2}
\end{figure}

Observing that the agents typically exhibit consensus behavior but it mismatches with the DeGroot prediction in \eqref{eq:limit_consensus}, in this section, we further investigate whether LLMs have an inherent and topic-dependent bias. 
To that end, we run a total of $150$ experiments on ``bitcoin'' and ``veganism'' subjects. 
We purposefully start these simulations with a high majority ``for'' or a high majority ``against'' initial opinion distribution, as demonstrated in the first column of Figure~\ref{fig:fig_4x2}. 
The corresponding final opinion distributions are given on the right column.

We see that in bitcoin-related experiments, the LLMs show an inherent negative bias: if they are initialized with against beliefs, they tend to stay like that, but if they are initialized with for beliefs, they can still change their opinions to against.
On the contrary, in veganism-related experiments, they have an inherent positive bias, as is evident from the final opinion distributions. Therefore, we find that the final opinion distributions show topic-dependent bias, even if these topics are also a part of the debate between humans. As other works suggested for factual-bias, the biases we observe can also be due to LLMs' preference and biases obtained during pretraining and RL-based alignment phases. 
Note that we observed cognitive biases in LLMs other than \texttt{Gemini} as well.

\subsection{Impact of the communication graph}
\label{impact_of_graph}

\begin{figure}[h]
    \centering
    \includegraphics[width=0.7\linewidth]{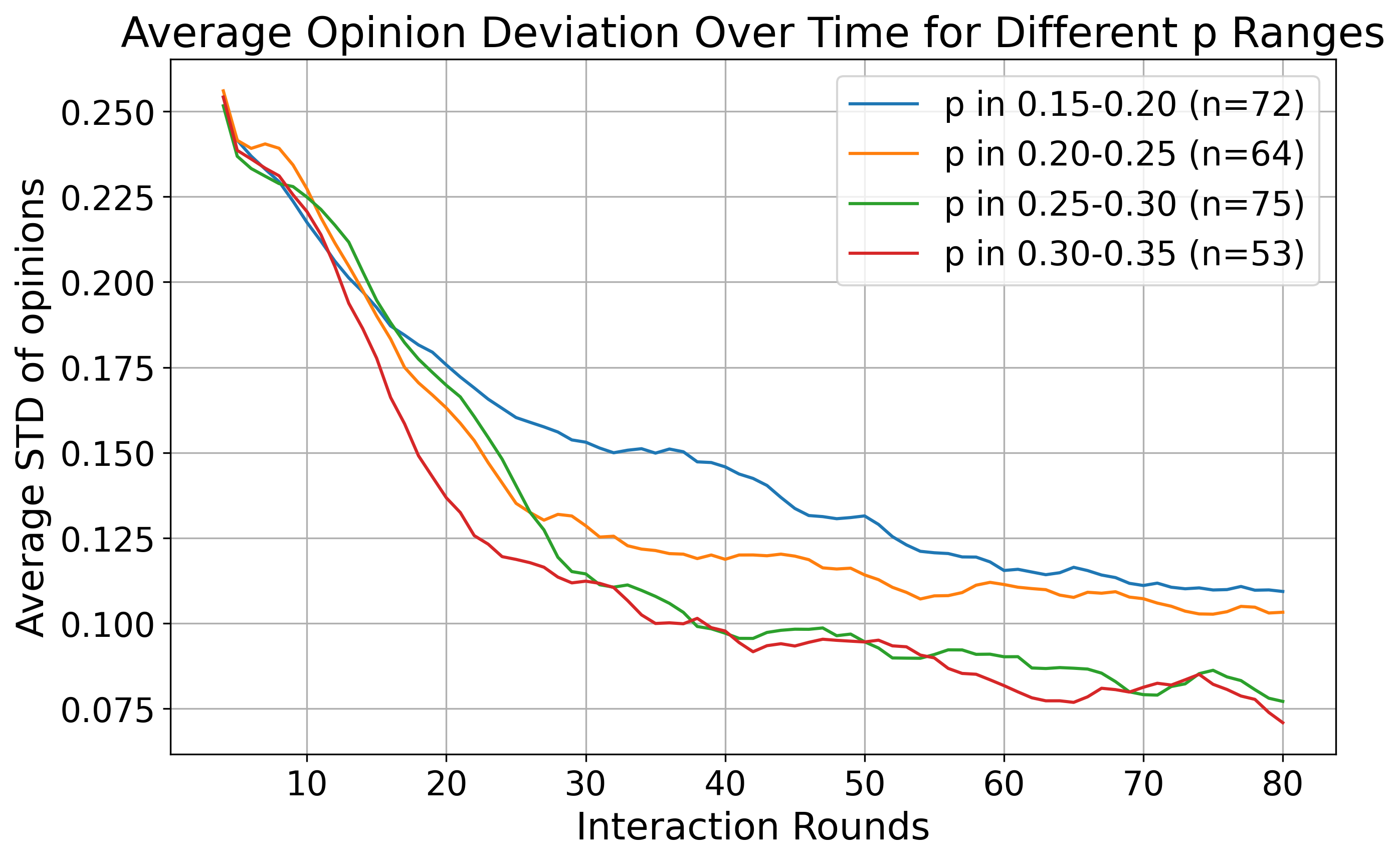}
    \caption{Average standard deviation of agents’ opinions over interaction rounds for different values of Erd\H{o}s--R\'enyi $p$. Each curve corresponds to a group of simulations within the indicated $p$ range, with $n$ denoting the number of experiments. Larger $p$ values result in faster convergence to the consensus with less disagreement.}
    \label{fig:fig_p}
\end{figure}

In this section, we analyze the effect of the communication topology on the opinion dynamics of LLMs. 
First, we evaluate the effect of the connectivity parameter $p$ of Erd\H{o}s--R\'enyi random graphs on the rate of convergence.
By definition, larger $p$ implies higher graph connectivity, and hence, we expect the convergence to be faster.
Figure~\ref{fig:fig_p} shows the disagreement between the opinions with respect to time, where each curve belongs to a specific interval of $p$-values.
We can see that higher $p$ associates with faster and stronger convergence, as expected.


\begin{figure}[h!]
    \centering
    \includegraphics[width=0.6\linewidth]{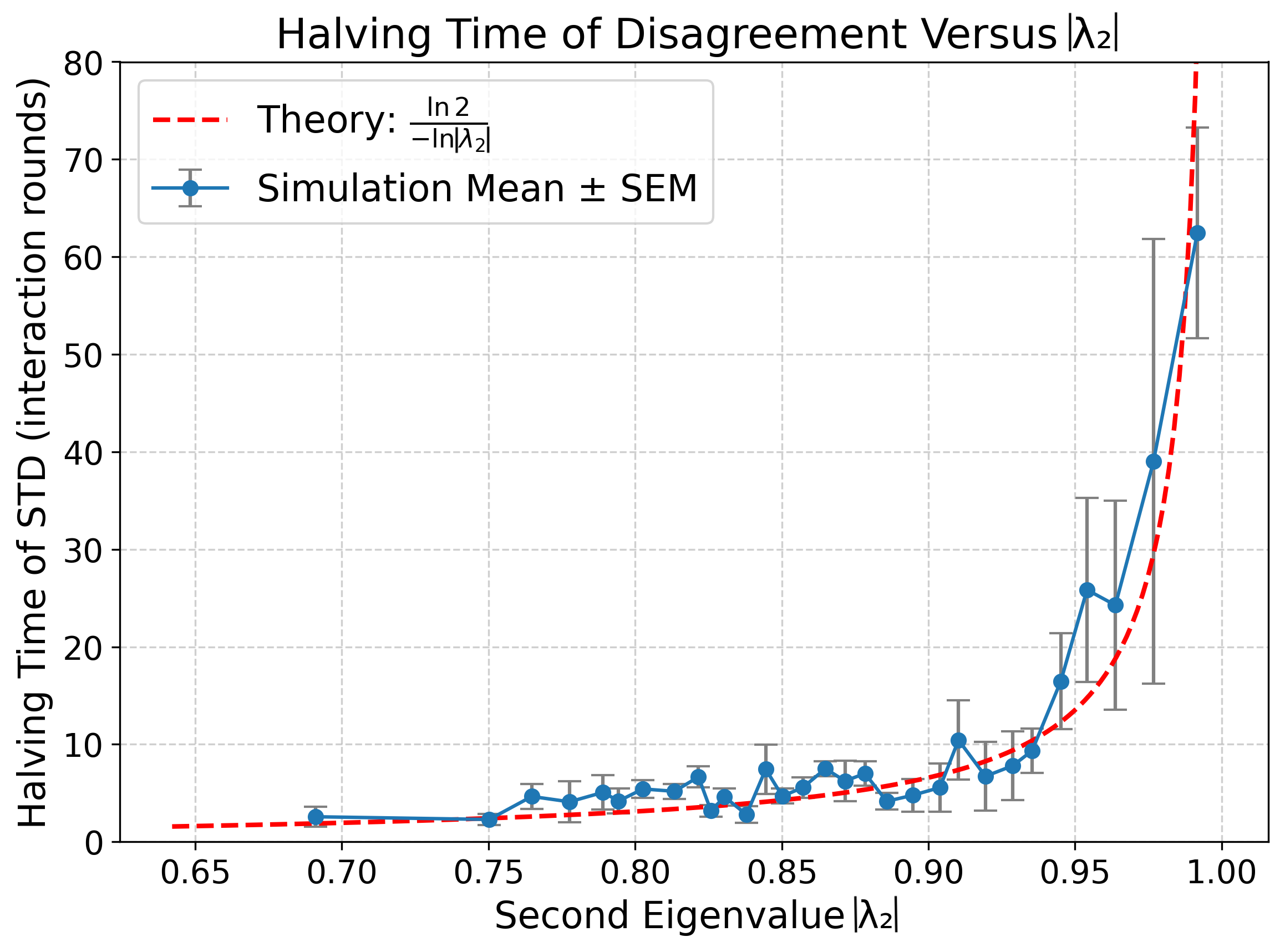}
    \caption{Halving time of disagreement between LLM agents, changing with respect to the second eigenvalue of the combination matrix. The eigenvalues are arranged into $30$ discrete bins, and the total number of experiments is $110$. The halving time shown on the y-axis is the mean across all experiments. SEM is denoted with bars around the mean. The dashed red curve shows the theoretical halving time.}
    \label{fig:fig_eig_halflife}
\end{figure}

Next, we turn our attention to the second-largest modulus eigenvalue $\lambda_2$ of $A$ as a measure of convergence rate, as explained in Section \ref{problem_formulation}.
The eigenvalue $\lambda_2$ is affected by the self-confidence weights of the agents in the combination matrix as well as different $p$ values.
Therefore, to have a wider range of $\lambda_2$ values across experiments, we conduct $113$ additional experiments with \textit{self-confident self-weight} also varying between $65\%$ and $90\%$.
According to \eqref{eq:horn_convergence}, the disagreement of opinions decrease exponentially, with $\lambda_2$ controlling the mixing rate.
In order to test whether this is also the case for LLM interactions, we compute the halving time of disagreement, that is, we compute the number of interaction rounds it takes for the standard deviation between discussing agents' opinions halves compared to the initial standard deviation. Following the theory described in Sec.~\ref{problem_formulation}, we expect this quantity to be proportional to \cite{xiao2003fast}:
\begin{equation}\label{eq:halving_formula}
    t_{\frac{1}{2}} \triangleq \dfrac{\ln2}{- \ln |\lambda_2|}.
\end{equation}
Figure \ref{fig:fig_eig_halflife} shows the mean of the empirical halving times as a function of $|\lambda_2|$, as well as the function in \eqref{eq:halving_formula}. 
The empirical results closely match the theory, which indicates that LLM-networks' convergence behavior follows the results from spectral graph theory.
In other words, although the Perron eigenvector–based average consensus in \eqref{eq:limit_consensus} does not align with the networked LLM simulations, the corresponding rate of convergence closely matches the experimental behavior.

\section{Concluding Remarks} 

In this work, we constructed a dataset of LLM agents interacting over randomized network topologies, across diverse topics and prompting strategies. Sentiment analysis of these interactions revealed that while the agents do not conform to the DeGroot consensus model, consensus still emerges as the deviation of opinions decreases exponentially through interaction rounds. This consensus is strongly influenced by inherent, topic-dependent biases acquired during pretraining, and its strength increases when interaction weights are imposed through system prompts. Furthermore, we observed that both the convergence rate and consensus strength grow with higher Erd\H{o}s--R\'enyi connectivity probabilities. The halving time of disagreement between agents was found to increase with the second-largest modulus eigenvalue of the graph combination matrix, and experimental results matched theoretical expectations based on spectral graph properties. Understanding the convergence rate of networked LLMs might be important for multi-agent system design, particularly in estimating the number of interaction rounds required for consensus under cost constraints. Though we qualitatively observed similar results with other LLMs than Gemini, further simulations could be done to explore other LLMs’ convergence behavior. Future work could also consider strategic agents with incentives or propaganda-sharing agents.

\clearpage

\appendix

\begin{center}
{\Large \textbf{APPENDIX}} 
\end{center}


\section{Dataset Organization}

This appendix describes the organization and directory structure of the
\texttt{Social-LLM-Networks} dataset, which is available on HuggingFace: https://huggingface.co/datasets/asl-epfl/Social-LLM-Networks.

The dataset consists of independent multi-agent experiments in which
networked LLM agents exchange opinions over a communication network. Each experiment is stored as a JSON file
and includes both the experimental parameters and the communication between agents. The dataset is organized hierarchically according to the LLM model used in the experiment (either \texttt{Gemini} or \texttt{OpenAI}), the experimental setting (main experiments or ablation studies), the discussion topic or the type of ablation. The overall structure of the repository is summarized in Figure \ref{fig:repo_structure}.

\begin{figure}[h]
\centering
\begin{tikzpicture}[
    box/.style={draw, rectangle, rounded corners, align=center,
                minimum width=2cm, minimum height=0.9cm},
    smallbox/.style={draw, rectangle, rounded corners, align=center,
                     minimum width=2.8cm, minimum height=0.8cm},
    arrow/.style={->, thick}
]

\node[box] (root) {Social-LLM-Networks};

\node[box] (gemini) [below left=1.4cm and 1.8cm of root] {gemini2flash};
\node[box] (gpt)    [below right=1.4cm and 1.8cm of root] {gpt5nano};

\draw[arrow] (root) -- (gemini);
\draw[arrow] (root) -- (gpt);

\node[smallbox] (main)     [below left=1.2cm and 0.6cm of gemini] {main};
\node[smallbox] (ablation) [below right=1.2cm and 0.6cm of gemini] {ablation};

\draw[arrow] (gemini) -- (main);
\draw[arrow] (gemini) -- (ablation);

\node[smallbox, text width=3cm] (topics)
[below=1.1cm of main]
{
--bitcoin \\ --euthanasia \\ --gene\_editing \\ ghosting \\
--remote\_work \\ --ronaldo \\ --vaping \\ --veganism
};

\draw[arrow] (main) -- (topics);

\node[smallbox, text width=3.7cm] (abl)
[below=1.1cm of ablation]
{
--biased\_start \\
--different\_selfweights \\
--weightless
};

\draw[arrow] (ablation) -- (abl);

\node[smallbox] (gptb) [below=1.2cm of gpt] {--biased\_start};
\draw[arrow] (gpt) -- (gptb);

\end{tikzpicture}
\caption{Organization of the \texttt{Social-LLM-Networks} repository.}
\label{fig:repo_structure}
\end{figure}

\subsection*{A.1 Directory}

\paragraph{Gemini2Flash:}
The \texttt{gemini2flash} directory contains experiments conducted using the
Gemini2Flash model. It is divided into two components:

\begin{itemize}
    \item \texttt{main/}: Contains the primary experimental runs, organized by
    discussion topic. Each topic subfolder (e.g., \texttt{bitcoin},
    \texttt{veganism}) contains JSON files corresponding to different experiments on that topic.
    
    \item \texttt{ablation/}: Contains 3 different types of studies:
    \begin{itemize}
        \item \texttt{biased\_start/}: Experiments with non-uniform
        initial opinion distributions. The majority of agents are initially for or initially against a given topic.
        \item \texttt{different\_selfweights/}: For each experiment, the self-confident agents' self-weight is randomly selected to be between 65-90\%.
        \item \texttt{weightless/}: Explicit self-weighting is removed from the prompts, but agents are still assigned characteristics (either open-minded or self-confident) through prompts.
    \end{itemize}
\end{itemize}

\paragraph{GPT5Nano:}
The \texttt{gpt5nano} directory contains experiments conducted using the
GPT5Nano model. This directory includes
a \texttt{biased\_start/} subfolder, which can be compared with the biased initialization
experiments also performed for \texttt{Gemini2Flash}.

\subsection*{A.2 JSON Experiment File Contents}

Each JSON file represents one experiment run and includes the following information:

\begin{itemize}

    \item Communication network topology
    \vspace{-0.8em}

    \item A chronological sequence of agent responses
    \vspace{-0.8em}

    \item All system prompts and initial prompts
    \vspace{-0.8em}

    \item Topic of discussion
    \vspace{-0.8em}

    \item Initial opinions of agents on the given topic
    \vspace{-0.8em}

    \item Stance scores between 0 and 1 for each response. (0 indicates Against while 1 indicates For.)
    \vspace{-0.8em}

    \item Graph type
    \vspace{-0.8em}

    \item Erdos Renyi $p$ value
    \vspace{-0.8em}

    \item Self-confident self-weight
    \vspace{-0.8em}

    \item Number of interaction rounds
    \vspace{-0.8em}

    \item Total execution time of the experiment
    \vspace{-0.8em}

    \item AI model
\end{itemize}

\clearpage

\section{Prompts}

The system prompts provided to agents after each interaction round are provided in Figure \ref{fig:system_prompts}.
The prompts which are used to assign initial opinions to agents are provided in Figure \ref{fig:initial_prompts}. \vspace{2em}

\begin{figure}[h]
\centering
\begin{tcolorbox}[
    width=\linewidth,
    colframe=blue,
    colback=white,
    title=System Prompts (Agent Personalities),
    fonttitle=\bfseries,
    breakable
]
\textbf{Self-Confident}

You are a person participating in a discussion about \{Bitcoin\}.
You are stubborn about your opinion and do not easily change it.
You are influenced by your own previous opinion with weight $0.80$
and by others' opinions with weight $0.20$.
You form your new opinion accordingly.

\vspace{1em}
\textbf{Open-Minded}

You are a person participating in a discussion about \{Bitcoin\}.
You are open-minded and frequently change your opinion based on others.
You take other participants' opinions for granted.
You are influenced by your own previous opinion with weight $0.60$
and by others' opinions with weight $0.40$.
You form your new opinion accordingly.

\end{tcolorbox}
\caption{System prompts defining agent characteristics and opinion update behavior.}
\label{fig:system_prompts}
\end{figure}

\begin{figure}[h]
\centering
\begin{tcolorbox}[
    width=\linewidth,
    colframe=orange,
    colback=white,
    fonttitle=\bfseries,
    title=Initial Opinion Prompts,
]

\textbf{Against}

I need you to impersonate someone who absolutely hates \{Bitcoin\}.
You are absolutely against and you hate \{Bitcoin\}. You HATE and detest \{Bitcoin\}.
Your initial opinion is: Absolutely against \{Bitcoin\}!
What do you think about \{Bitcoin\}. Keep your answer under 150 words. Do not exceed 150 words. State your opinion.

\vspace{1em}
\textbf{Neutral}

I need you to impersonate someone who is neutral towards \{Bitcoin\}.
You are neither for nor against \{Bitcoin\}. You are completely neutral.
Your initial opinion is: absolutely neutral towards \{Bitcoin\}.
What do you think about \{Bitcoin\}.
Keep your answer under 150 words. Do not exceed 150 words. State your opinion.

\vspace{1em}
\textbf{For}

I need you to impersonate a true \{Bitcoin\} lover. You are a huge lover of \{Bitcoin\}.
You absolutely LOVE \{Bitcoin\} and you are very very in favor of \{Bitcoin\}.
You LOVE \{Bitcoin\}. Act like you are a huge fan! Your initial opinion: Absolutely for \{Bitcoin\}.
What do you think about \{Bitcoin\}. Keep your answer under 150 words.
Do not exceed 150 words. State your opinion.

\end{tcolorbox}
\caption{Initial system prompts used to assign initial opinions to agents for the discussion topic of \{Bitcoin\}.}
\label{fig:initial_prompts}
\end{figure}

\clearpage
\section*{Acknowledgment}
This work was performed while Iris Yazici was a visiting student at EPFL. The work of Mert Kayaalp was partially supported by UBS Switzerland AG and its affiliates through UBS-IDSIA AI Lab.

\bibliographystyle{IEEEbib}
\bibliography{strings,refs}

\end{document}